# Critical properties of superconducting $Ba_{1-x}K_xFe_2As_2$

A.Bharathi, Shilpam Sharma, S. Paulraj, A. T. Satya, Y. Hariharan and C.S.Sundar

Materials Science Division, Indira Gandhi Centre for Atomic Research, Kalpakkam 603102, India

Magnetisation and magnetoresistance measurements have been carried out on superconducting $Ba_{1-x}K_xFe_2As_2$ samples with x=0.40 and 0.50. From low field magnetization data carried out at different temperatures below $T_C$, $H_{C1}$ has been extracted. The plot of $H_{C1}$ versus temperature shows an anomalous increase at low temperatures. From high field magnetization hysterisis measurements carried out in fields up to 16 T at 4.2 K and 20 K, the critical current density has been evaluated using the Bean critical state model. The $J_C$ determined from the high field data is $>10^4 A/cm^2$ at 4.2 K and 5 T. The superconducting transitions were also measured resistively in increasing applied magnetic fields up to 12 Tesla. From the variation of the $T_C$ onset with applied field, $dH_{C2}/dT$ at $T_C$ was obtained to be -7.708 T/K and -5.57 T/K in the samples with x=0.40 and 0.50.

Corresponding Author
A.Bharathi
Materials Science Division
Indira Gandhi Centre for Atomic Research
Kalpakkam. 603102. India
bharathi@igcar.gov.in

# INTRODUCTION

The recent discovery of superconductivity in Fe based arsenides [1] has elicited enormous excitement amongst condensed matter physicists, since many of the properties in these compounds bear a strong resemblance to the much investigated high temperature superconducting oxides. Several classes of compounds have been discovered systematically in the last few months [1,2,3,4]. The two major families of FeAs superconductors that have been investigated are REOFeAs systems, broadly termed 1111 compounds [1], where RE is an early rare-earth and the two layered $MFe_2As_2$ systems, termed as the 122 compounds[2], where M stands for an alkaline earth, Ca,Sr,Ba or the divalent rare earth Eu. Hole and electron doping in the REO layers of the 1111 compounds induce the superconducting state at the expense of the spin density wave (SDW) ground state present in the pristine sample [5]. Likewise, K substitution at the M site [2] and Co substitution at the Fe [6] site in the $MFe_2As_2$ compounds, introduce holes/electrons respectively in the system giving rise to superconductivity, destroying the SDW state [7]. Of the two classes of compounds, the 122 compounds have been more thoroughly investigated, since single crystals were available form a very early stage [8]. Like the cuprates these compounds show an increase in the superconducting transition temperature with increase in external pressure, if optimally doped [9]. However, in contrast to cuprates, it is seen that the parent compound can be self doped with the application of moderate pressure: For example, in the $CaFe_2As_2$ system, application of 1.5 GPa results in the disappearance of the SDW state and the occurrence of superconductivity at 19 K[10]. Similar pressure induced superconductivity is also observed in the $BaFe_2As_2$ and $SrFe_2As_2$ systems, which however is controversial [11]. Much of the early literature in this compound was devoted to understanding the SDW transition and the associated structural changes in the pristine compound and its disappearance with chemical substitution [12].

In a short span of time, there have been several studies on the characterization of the superconducting state of the arsenide superconductors. These include the measurement of critical fields, superconducting gaps, penetration depth, coherence lengths etc[13-24]. Many of these experiments have been carried out on single crystalline samples, to investigate the anisotropic properties[14-17]. While there are discrepancies in the reported values of coherence lengths, penetration depths etc, the consensus arrived at indicates that the arsenide system is a type-II superconductor with a Ginzburg-Landau parameter, κ to be ~80[17]. A variety of experiments have pointed to the existence of two superconducting gaps[18,19,20,21], as is also supported by electronic structure calculations[22]. The anisotropy in the physical properties of 122 arsenide appears to be much lesser than in the cuprate superconductor, suggesting that it is a 3D superconductor[14,15,16,17,24]. The proximity to SDW transition, existence of two superconducting gaps, and 3D nature of superconductivity have to form the crucial underpinnings of the theoretical models of superconductivity in this system.

In this paper we present the critical properties of the $Ba_{1-x}K_xFe_2As_2$ superconductor (x=0.40 and x=0.50) in fields up to 16 Tesla. We employ low field magnetization measurements to obtain lower critical fields $H_{C1}$ as a function of temperature and magneto-resistance measurements up to 12 Tesla to obtain an estimate of the upper critical field $H_{C2}(0)$. The high field magnetization measurements upto 16 tesla were used to obtain the critical current density at 4.2 K and 20K. The experiments have been carried out in two samples, that have been prepared in two different ways (see below) resulting in K concentrations, x=0.40 and x=0.50.

## EXPERIMENTAL DETAILS

Towards a controlled synthesis of these arsenide samples we employ a stainless steel chamber that can be pressure locked with 50 bar of pure Ar gas, to prevent volatisation of As. This procedure dispenses off with the use of sealing in quartz tubes and this we believe is a safer way to synthesise these arsenides. We have followed two recipes for synthesis: In the first method (sample 1) we first synthesise FeAs by heat treating an intimate mixture Fe powder (99.999%) and As powder (99.9999%) at 600 C for six hours under pressure. Stoichiometric quantities of Ba, K and FeAs of composition $Ba_{1-x}K_xFe_2As_2$ with nominal x=0.45 are then weighed and assembled together into Ta crucibles in a He/Ar containing dry box. This is assembled into the stainless steel tube and filled with 50 bar of ultra pure Ar and ramped at 50 C/hour to 800 C followed by soaking at that temperature for 10 hours. The product is then ground and pelletised in the dry box and subsequently sintered at 950 C for 5 hours. In the second method (sample 2) we skipped the FeAs step and started with a mixture of Fe and As powders and Ba and K chunks. This was heat treated at 800 C for 4 hours followed by pelletising and sintering at 950 C for 5 hours.

The X-ray characterisation of the samples were done in a STOE diffractometer operating in the Bragg-Brentano geometry, indicated the formation of the I4/mmm, $ThCr_2Si_2$ structure. A small amount of unreacted FeAs was seen in sample 1, whereas in sample 2 in addition to FeAs some $Ba_3As_{14}$ was also present [25]. Using the STOE program the lattice parameters were obtained to be a=0.39159 nm and c=1.3318 nm in sample 1 and a=0.3899 nm and c=1.3411 nm in sample 2 respectively. Using these lattice constants and the experimental variation of a,c lattice parameters with K concentration[26] the K contents were estimated to be x~ 0.40 and x~ 0.50 in sample 1 and sample 2.

The resistivity versus temperature was measured in the four probe geometry, and ac susceptibility measurements were carried out using a mutual inductance set-up, in a dipper cryostat. The magnetoresistance measurements were carried out in the four probe geometry in an apparatus with field capability of up to 12 Tesla operable in the 6 K to 300 K temperature range. The magnetization measurements were carried out in a CRYOGENIC make, liquid helium based, vibrating sample magnetometer operating at 20.4 Hz capable of a magnetic field range of up to 16 Tesla and temperature variation from 2 K to 300 K.

## RESULTS & DISCUSSIONS

Fig.1 shows the resistivity measured in the sample with x=0.40 in the 4 K to 300 K temperature range. The room temperature resistivity is 2.2 mΩcm, and the sharp superconducting transition is clearly seen. The inset (b) of Fig.1 shows the diamagnetic signal in the same sample. The onset of the superconductivity occurs at 37.5 K with a $\Delta T_C$ of 3 K (obtained from the temperature difference between onset and downset of transition). The second sample that has a K fraction of x=0.50 has a room temperature resistivity of 1.2 mΩcm and the $T_C$ onset determined by ac susceptibility is 35.7 K with $\Delta T_C$ of 10 K. The normal state resistivity shown in Fig.1, does not exhibit the linear temperature dependence seen in High $T_C$ cuprates or the Bloch-Gruneisen behaviour seen in $MgB_2$[27]. Instead the temperature dependence of resistivity in the normal state varies as $T^2$ upto 70 K, beyond which it becomes sublinear in T, very similar to that seen in A15 compounds [28]. The residual resistance ratio (RRR) is large at ~8, similar to that obtained by Rotter et. al[26].

The resistance measurements carried out in the 5 K to 45 K temperature for various external magnetic fields up to 12 Tesla are shown in Fig.2 for the x=0.40 sample. It is clear from the figure that the application of magnetic field does not broaden the superconducting transitions in the $Ba_{1-x}K_xFe_2As_2$ samples, unlike that seen in the oxyarsenides [13], implying that there is not much granularity or anisotropy in the double FeAs system. The lack of significant anisotropy has been amply demonstrated in on single crystals of the Ba122 compounds[14-16]. Further it is also clear from the figure that there is a small positive magnetoresistance in the normal state of the sample which is ~3.7%. The plots of the $T_C$ onsets (defined as shown in Fig.2), as a function of applied external field are displayed in Fig.3a and 3b for x=0.40 and x=0.50 respectively. The linear fit of the $H_{C2}$ versus T curve gives a value of $dH_{C2}/dT$ at $T_C$ in x=0.40 sample to be -7.708 T/K and in the sample with x=0.50 to be -5.57 T/K. Based on WHH formula[29] in the dirty limit, this results in an evaluation of $H_{C2}(0)$ in the two samples to be 202 T and 147 T respectively which is much larger than the Chandrasekhar-Clogston paramagnetic limit [30] of ~70 T for this compound.

In the dirty limit, $(dH_{C2}/dT)_{Tc}$ is proportional to the normal state resistivity and linear coefficient of specific heat[29]. The lower values of the normal state resistivity and $T_C$ in the x=0.50 sample may be responsible for the lower $(dH_{C2}/dT)_{Tc}$ observed in this sample as compared to that with x=0.40. These values of $(dH_{C2}/dT)_{Tc}$ seen in Fig.3 are similar to that observed perpendicular to the c axis in single crystals of $Ba_{1-x}K_xFe_2As_2$ [9]. It should be noted that these large $(dH_{C2}/dT)_{Tc}$ values are generic to the arsenide compounds and are particularly large when compared to that observed in High $T_C$ superconductors and C substituted $MgB_2$[31]. Recent high field measurements seem to suggest that $H_{C2}(0)$ in these systems is in the paramagnetic limited regime [32].

In order to obtain $H_{C1}$, low field M versus H curves were measured by changing the magnetic field in steps of 0.005 T from 0 T to 0.4 T in the forward cycle. Care was taken to ensure that before each such M versus H measurement, the sample was taken to the normal state. These curves measured at several temperatures are shown in Fig.4 for the x=0.4 sample. It is clear from the figure that with decrease in temperature the linear part corresponding to the Meissner state occurs up to larger magnetic fields. The field point at which the M versus H curve deviates from linearity (as shown in the inset of Fig.4) is identified as $H_{C1}$ at that temperature. Similar sets of measurements have also been carried out in the sample with x=0.50. $H_{C1}$ was evaluated from the experimental data as a function of temperature in both the samples. A plot $H_{C1}$ versus $T/T_C$ in the two samples is shown in Fig.5. The Ginzburg Landau temperature dependence of $H_{C1}$, given by $H_{C1}= \ln(\kappa)\phi_0/(4\pi\lambda^2)$ with a temperature dependence of $\lambda=\lambda(0)[1-(T/T_C)^4]^{(-1/2)}$ expected for a single gap superconductor [33,34] is shown as a solid line in Fig.5. An unusual increase of $H_{C1}$ at temperatures below 0.4 $T_C$, overriding on the Ginzburg-Landau behaviour is clearly evident from Fig.5 for the x=0.40 sample. The anamolous increase occurs at a lower temperature ~0.25$T_C$ in the sample with x=0.50. What is striking about the figure is that the variation of $H_{C1}$ with temperature in the two samples is identical at high temperatures. Yet another feature worth noting from Fig.5 is that the low temperature upturn seen in the experimental data for both the samples has a nearly linear dependence on temperature.

The low temperature deviation from expected temperature dependence of $H_{C1}$ is seen in both samples but at different temperatures, probably because of their differing K contents. Recent measurements of $H_{C1}$ done on single crystals of $Ba_{0.55}K_{0.45}Fe_2As_2$, using Hall array detectors show a similar upturn in $H_{C1}$ at low temperatures[35]. Our results are slightly different from that seen in single crystals [35], in that the $H_{C1}$ does not saturate at low temperatures. A behaviour similar to that seen in Fig.5 of our data has been observed in recent measurements on polycrystalline F doped oxyarsenide compounds[36,37]. The linear T dependence of the superfluid density (proportional to $H_{C1}$) at low temperatures, in high $T_C$ cuprates has been understood to arise due to nodes in the superconducting gap[38]. Our data on $H_{C1}$ shown in Fig.5 therefore seems to suggest that the $Ba_{1-x}K_xFe_2As_2$ system has two superconducting gaps and the lower of the two gaps has nodes and is unconventional.

The magnetization hysterisis curves were measured in the two samples at 4.2 K and 20 K. For this measurement the magnetic field was ramped from 0 T to 16 T in the forward direction and then ramped to 16 T in the reverse direction and then back to 0 T. The measured magnetization in the entire forward and reverse cycling of magnetic field at 4.2 K is shown in Fig.6 for the x=0.40 sample. It is clear from the figure that the hysteretic behaviour persists up to a large field of 16 T at 4.2 K. It is also clear from Fig.6 that the hysteristic curve is superposed on a ferromagnetic background, possibly arising from a small amount of unreacted FeAs present in the sample[39]. A much larger ferromagnetic background was seen in polycrystalline samples in the oxyarsenides [23]. Using the hysteretic magnetization data, $J_C$ is evaluated using the Bean critical state model, $J_C=2(M_+-M_-)/d$, where $M_+$ and $M_-$ are the measured magnetization in the forward and reverse cycle at a given magnetic field and d is the diameter of the grains. With an average grain size of 10 μm obtained from scanning electron microscopy, $J_C$ is evaluated at 4.2 K and 20 K for x=0.40 and x=0.50, which are shown in Fig.7a and Fig.7b respectively. It is clear from the figure that at 4.2 K, $J_C$ is larger than $10^4$ A/cm$^2$ even at 12 T. Such large values have also been observed in the oxyarsenides at 5 K[23]. These $J_C$ values are however much larger than that observed in specifically processed C doped MgB$_2$ [31].

## SUMMARY

In conclusion we have measured the critical properties of superconducting Ba$_{1-x}$K$_x$Fe$_2$As$_2$. We find an anomalous increase in H$_{C1}$ at low temperature that agrees with recent measurements on single crystals of this compound, suggesting the presence of more than one gap. The dH$_{C2}$/dT and the critical current obtained from magnetization measurements in this compound are notably high.

39. Magnetic hysterisis curve measured on FeAs powder up to 16 T indicates that it is a very soft ferromagnet with a negligible remnance.

**FIGURE CAPTIONS**

**Fig.1** The variation of resistivity with temperature in $Ba_{1-x}K_xFe_2As_2$ for x=0.40. Inset a. shows the structure of the unit cell [7] and (b) shows the diamagnetic signal in the same sample.

**Fig.2** The resistance measured as function of temperature for sample 1 is shown for three field 0 T, 6 T and 12 T, measurements at all fields are not shown for clarity. The dotted line shows how the $T_C$ onset was determined for the 12 T data.

**Fig.3**. The variation of the $T_C$ onset obtained from R(T) measurements with the application of magnetic field for $Ba_{1-x}K_xFe_2As_2$ (a) x=0.40 and (b) x=0.50, the solid line in each of the panels is the linear fit used to obtain $dH_{C2}/dT$ at $T_C$.

**Fig.4** The variation of M versus applied field in the low field range of up to 0.4 T, shown for temperatures indicated for $Ba_{1-x}K_xFe_2As_2$, x=0.40. Inset shows how $H_{C1}$ is extracted from measured data at 5 K.

**Fig.5**. The variation in $H_{C1}$ versus the normalized temperature for $Ba_{1-x}K_xFe_2As_2$, x=0.40 and x=0.50. The anomalous increase in $H_{C1}$ at low temperature can be easily discerned. The solid line is the temperature variation of $H_{C1}$ expected from Ginzburg-Landau expression as explained in text.

**Fig.6** Variation of magnetization as a function magnetic field (0 to 16 T to -16 T and back 0 T) showing the magnetic hysterisis in $Ba_{1-x}K_xFe_2As_2$, x=0.40, at 4.2 K.

**Fig.7** The variation the critical current $J_C$ obtained from magnetization hysterisis loops at 4.2 K and 20 K in $Ba_{1-x}K_xFe_2As_2$ (a) x=0.40(b) x=0.50.

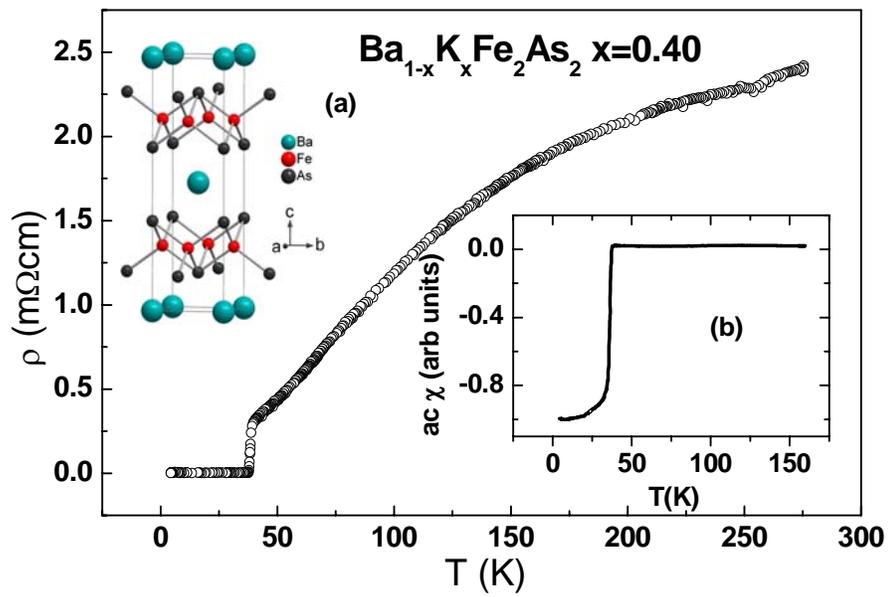

Fig.1

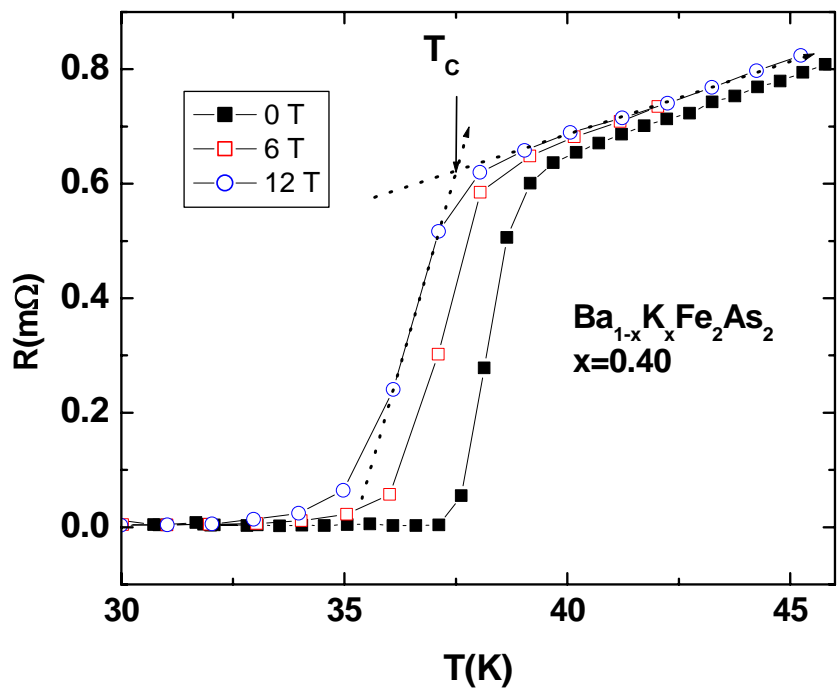

Fig.2

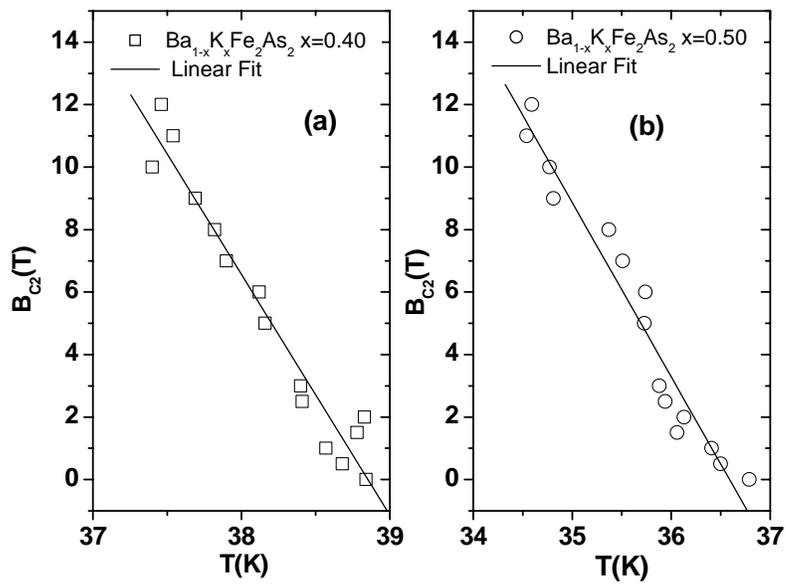

Fig.3

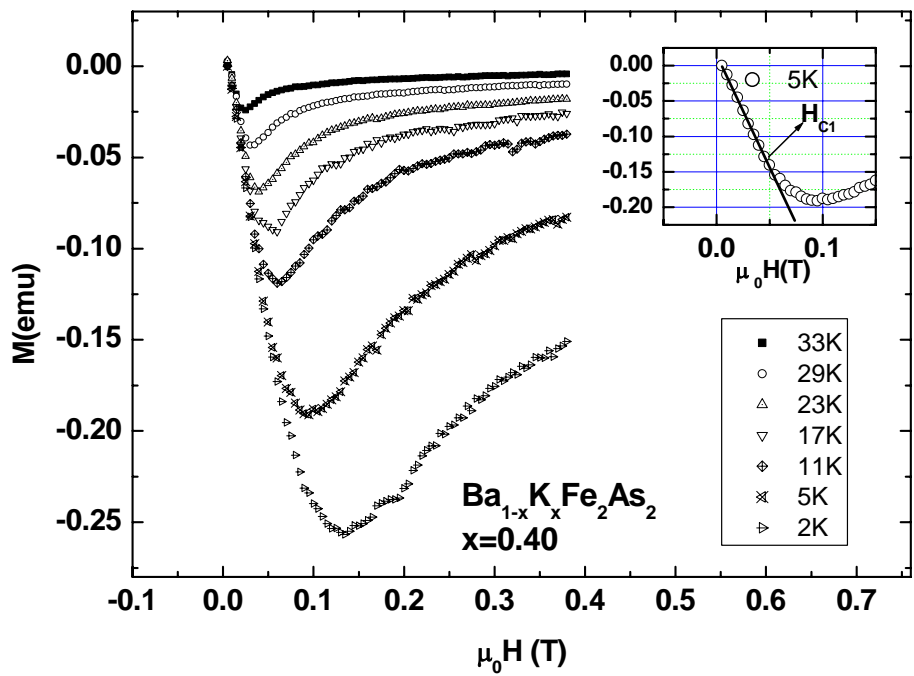

Fig.4

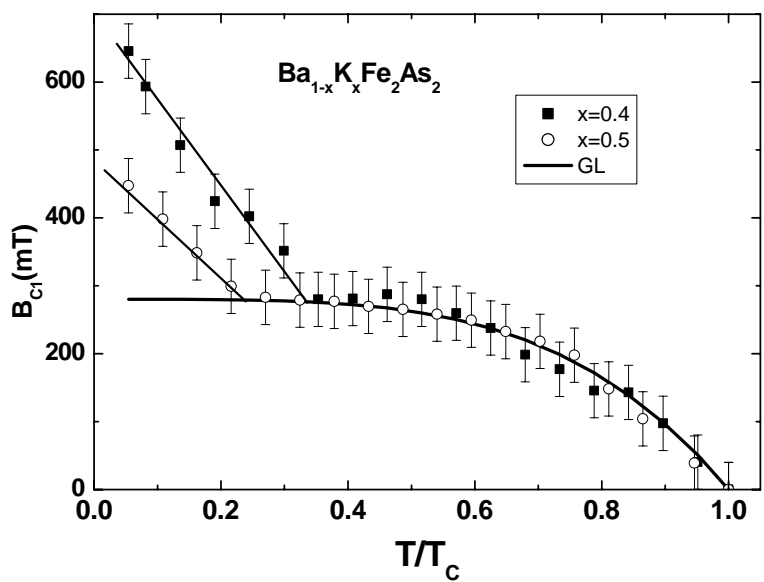

Fig.5

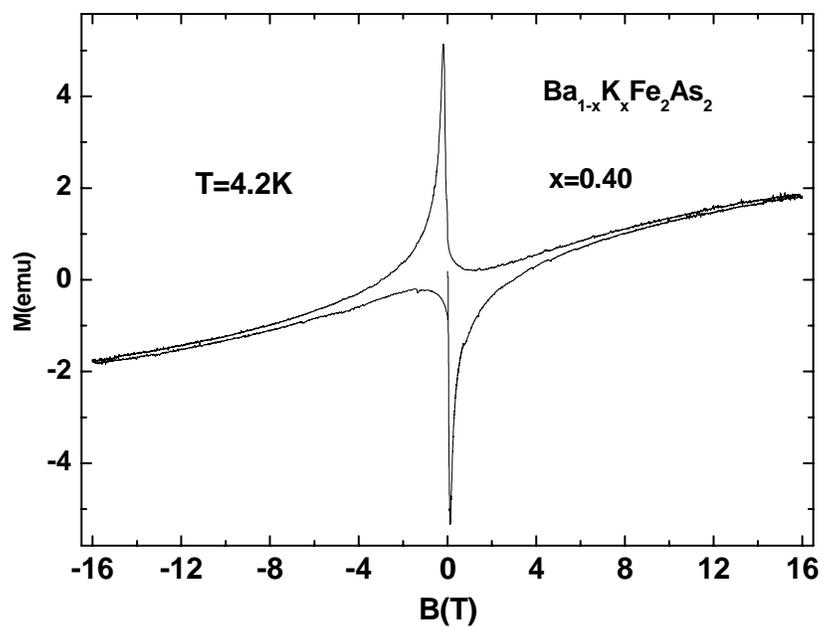

Fig.6

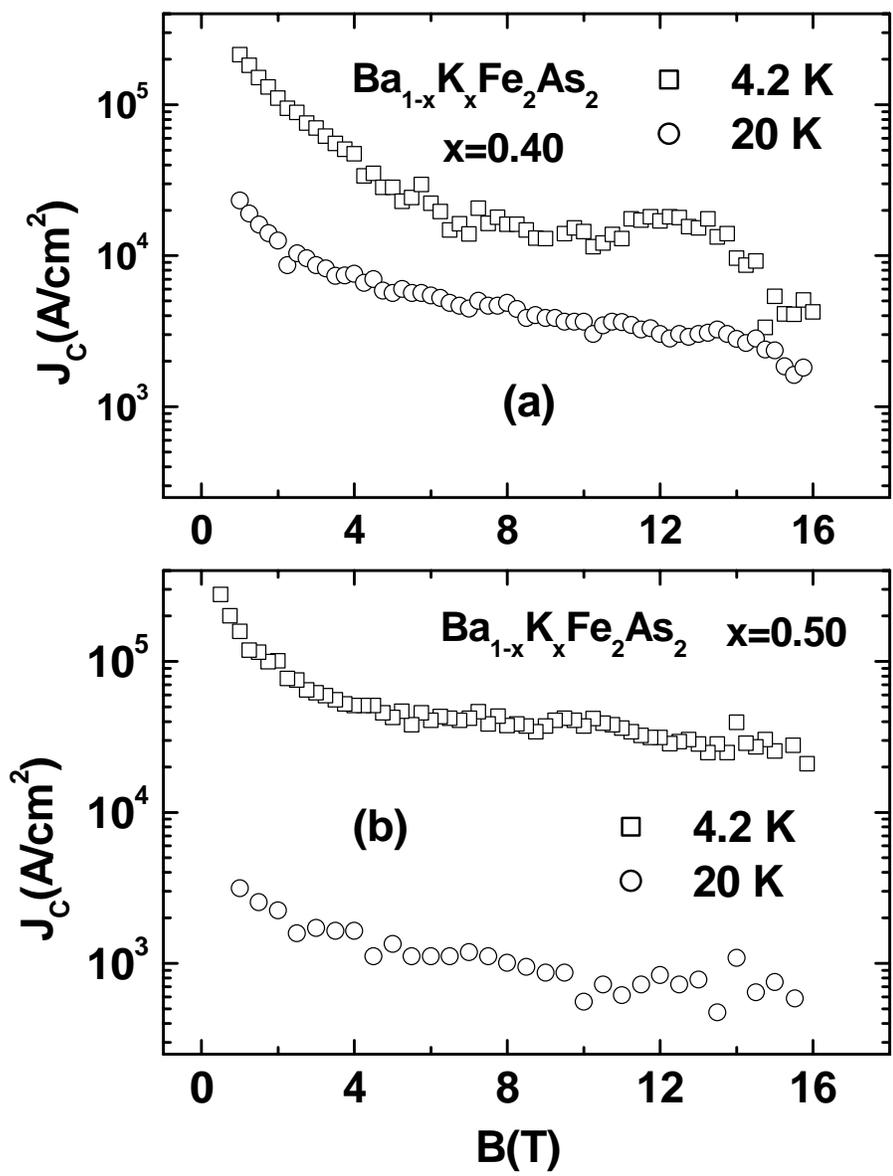

Fig.7